\documentclass[aps,prd,preprint,superscriptaddress,showpacs]{revtex4}

\usepackage{graphicx}
\usepackage{amsmath}
\usepackage{bm}

\def\ee{\end{eqnarray}}

\def\=:{=\hspace{-.7em}\raisebox{1.1ex}{.}\hspace{.1em}\raisebox{-0.2ex}{.} }

\def\ee{\end{eqnarray}}

\def\=:{=\hspace{-.7em}\raisebox{1.1ex}{.}\hspace{.1em}\raisebox{-0.2ex}{.} }

\newcommand {\beq}{\begin{eqnarray}}
\newcommand {\eeq}{\end{eqnarray}}
\newcommand {\non}{\nonumber\\}

\newcommand {\1}[1]{\frac{1}{#1}}

\newcommand {\ph}{\varphi}

\newcommand {\del}{\partial}

\newcommand {\tr}{{\rm tr}\,}

\begin{document}


\title{Instantons confined by monopole strings
}


\author{Muneto Nitta}

\affiliation{
Department of Physics, and Research and Education Center for Natural 
Sciences, Keio University, Hiyoshi 4-1-1, Yokohama, Kanagawa 223-8521, Japan\\
}


\date{\today}
\begin{abstract}
It is known that monopoles can be confined by vortex-strings in $d=3+1$
while vortices can be confined by domain-lines in $d=2+1$.
Here, as a higher dimensional generalization of these, 
we show that Yang-Mills instantons can be confined by monopole-strings 
in $d=4+1$.
We achieve this by putting the system into the Higgs phase in which the configuration can be constructed inside a non-Abelian vortex sheet.

\end{abstract}

\maketitle

\section{Introduction}
\begin{figure}[ht]
\begin{center}
\begin{tabular}{ccc}
$d=2+1$ & 
\includegraphics[width=0.3\linewidth,keepaspectratio]{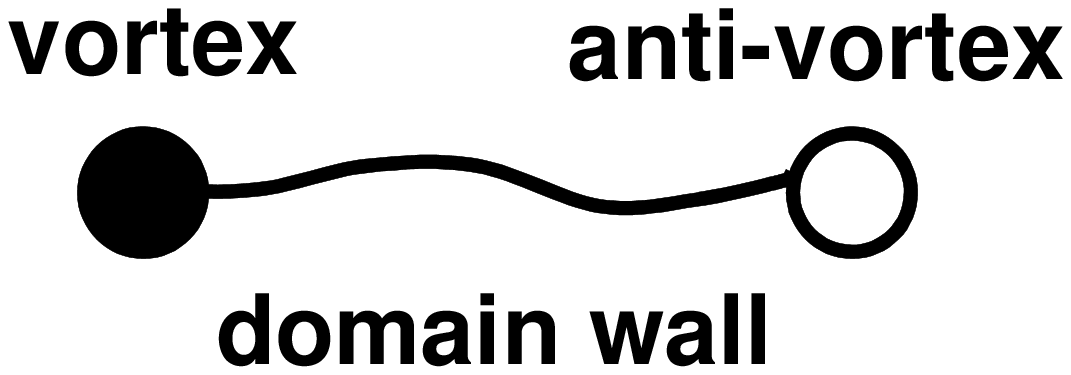}&
\includegraphics[width=0.3\linewidth,keepaspectratio]{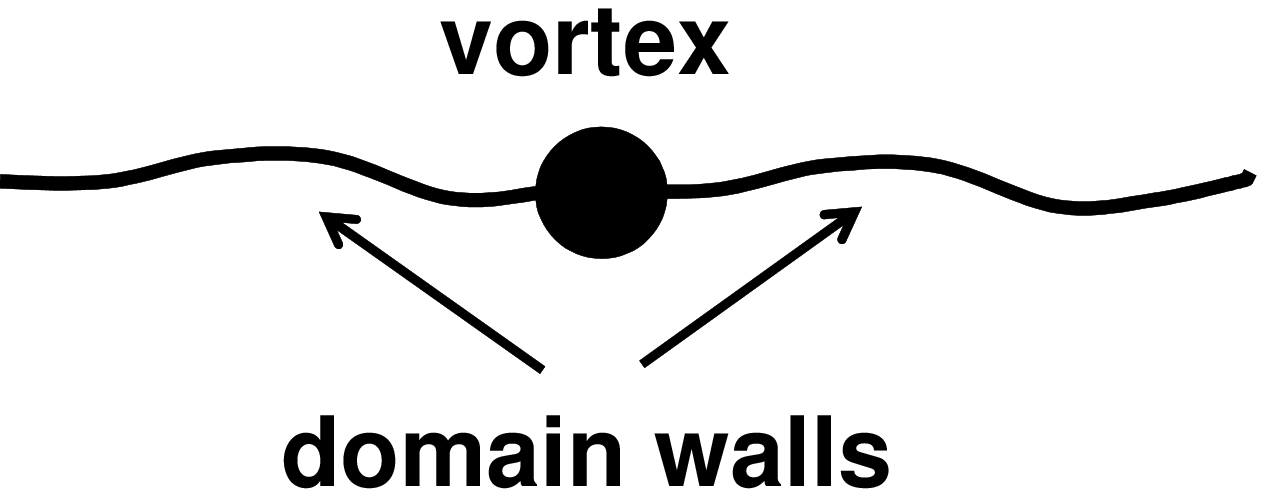}\\
&(a) & (b)\\
$d=3+1$ &
\includegraphics[width=0.3\linewidth,keepaspectratio]{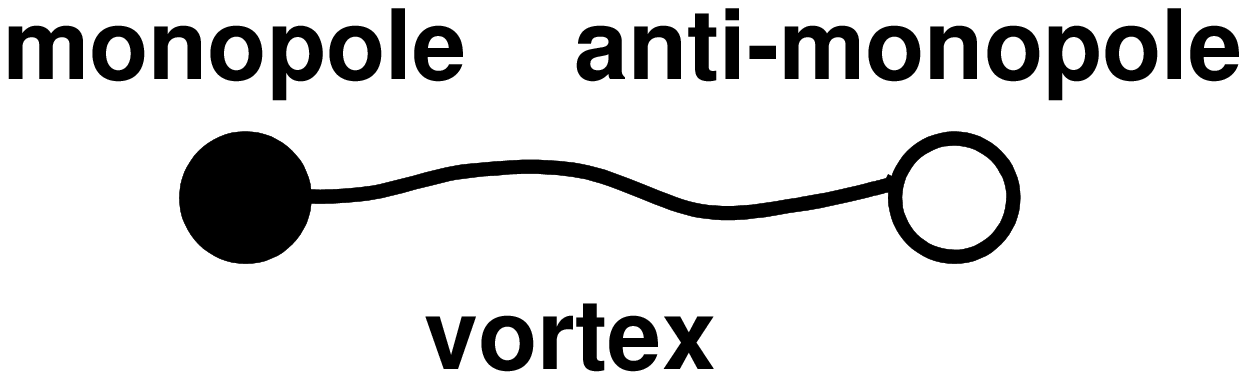}&
\includegraphics[width=0.3\linewidth,keepaspectratio]{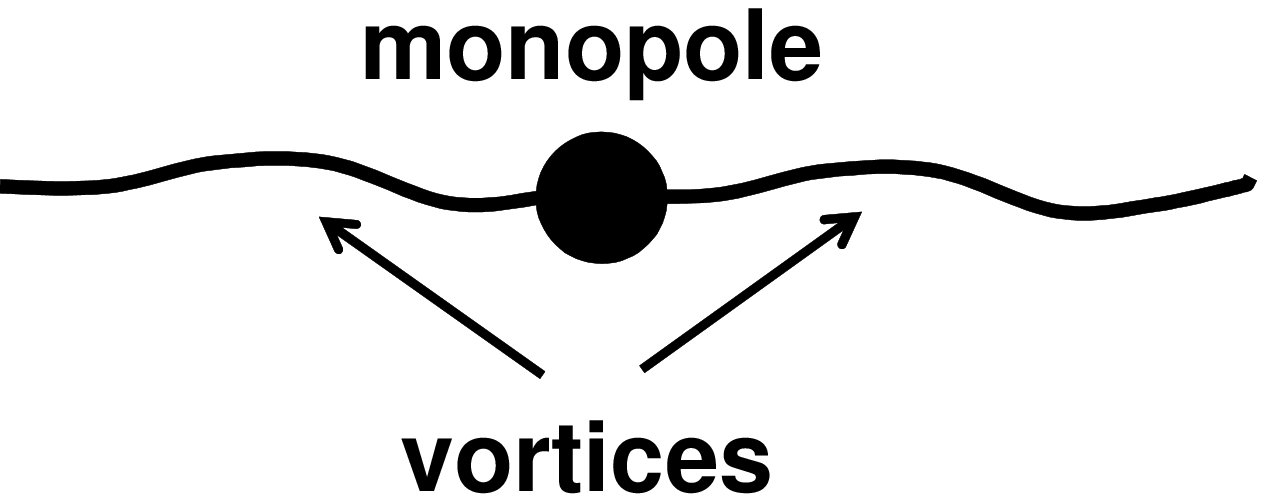}\\
&(c) & (d)\\
$d=4+1$ &
\includegraphics[width=0.3\linewidth,keepaspectratio]{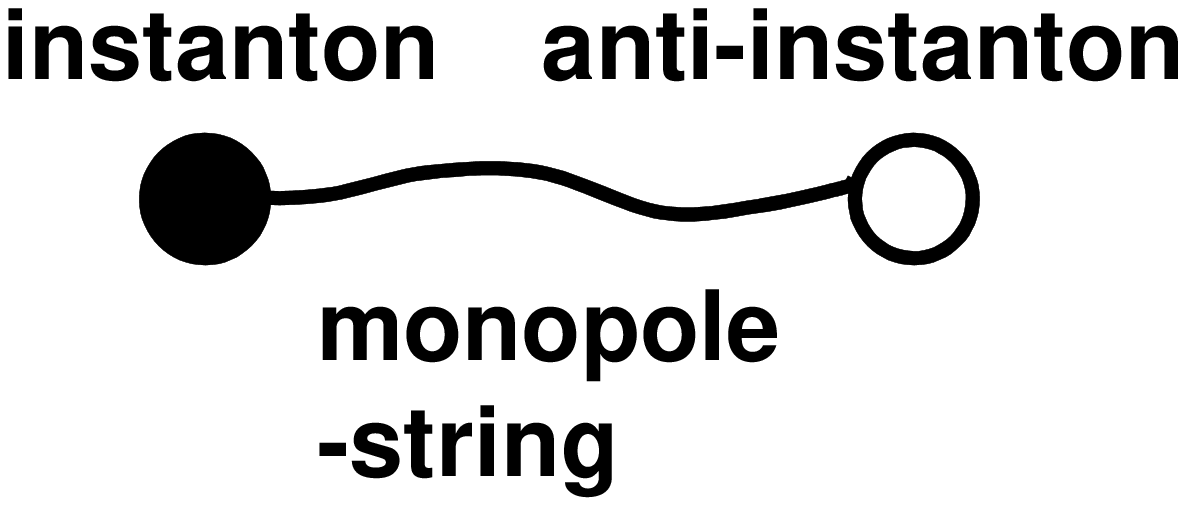}&
\includegraphics[width=0.3\linewidth,keepaspectratio]{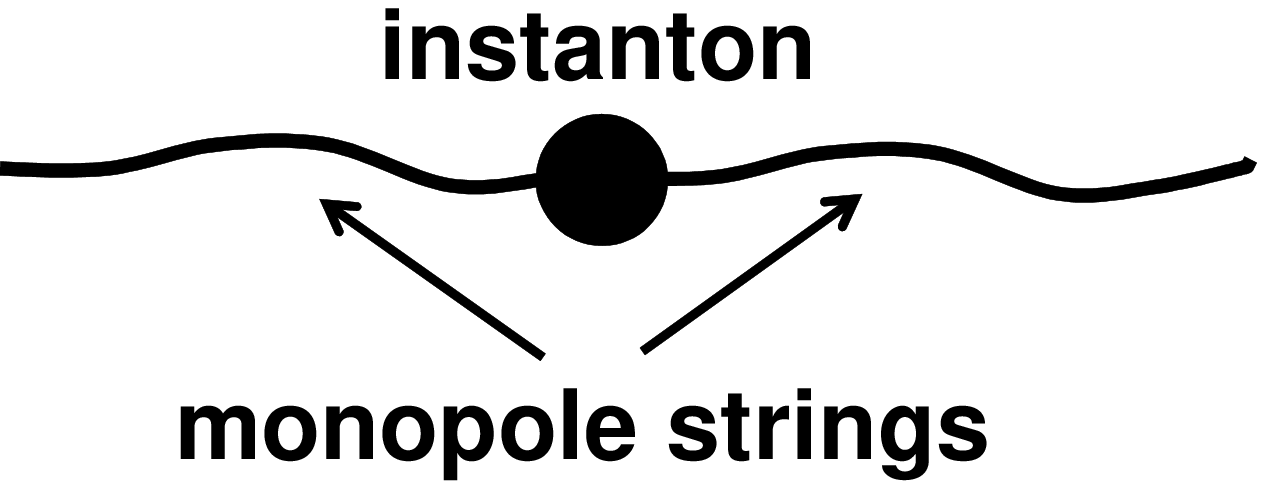}\\
&(e) & (f)
\end{tabular}
\caption{Soliton confinements. (a) Unstable and (b) stable vortex confinements by domain lines, (c) unstable and (d) stable monopole confinements by vortex-strings, and (e) unstable and (f) stable instanton confinements by monopole-strings. \label{fig:confined-defects}}
\end{center}
\end{figure}

Quark confinement is one of the most challenging problems 
in modern high energy physics. 
Color electric fluxes are expelled or squeezed in the QCD vacuum, 
and quarks are confined by electric fluxes. 
A quark and an anti-quark are connected by a color electric flux, 
constituting a meson bound state. 
This configuration shows a linear potential with respect to the distance 
between a well-separated quark and anti-quark \cite{Bali:2000gf}.  
This can be understood in a dual superconductor picture, 
in which magnetic fluxes are expelled or squeezed in dual Higgs vacuum, 
as in conventional superconductors. 
In the Higgs vacuum where the $U(1)$ gauge symmetry is broken, 
a magnetic monopole and an anti-monopole are confined by a magnetic flux, 
that is, a magnetic vortex [Fig.~\ref{fig:confined-defects}(c)] \cite{Nambu:1974zg}.
In the standard model of the electroweak unification, 
the electroweak $Z$-string ends on a Nambu monopole \cite{Nambu:1977ag,Eto:2012}.
In these models, a long vortex is merely metastable and can decay through quantum tunneling by creating a pair of a monopole and an anti-monopole \cite{Preskill:1992ck}. 
These configurations were also studied in cosmology 
to resolve the monopole problem 
\cite{Langacker:1980kd,Vilenkin:2000};
monopoles and anti-monopoles annihilate in pairs pulled by the tension of a vortex-string connecting them, if the $U(1)$ gauge symmetry is spontaneously broken in the Higgs phase  at some stage in the early Universe.
Recently, these configurations have been generalized to a non-Abelian monopole 
confined by non-Abelian vortices to discuss non-Abelian duality 
in supersymmetric QCD \cite{Auzzi:2003fs}. 
On the other hand, 
stable bound states of a magnetic monopole and vortices have been found in supersymmetric QCD \cite{Tong:2003pz,Eto:2004rz,Eto:2006pg,Shifman:2007ce,Nitta:2010nd}. 
In this case, one magnetic monopole is attached by two vortices from both of its sides [Fig.~\ref{fig:confined-defects}(d)]. 
When the tensions of the both vortices are equal, the total configuration is stable. This is indeed the case of supersymmetry 
since the vortices are Bogomol'nyi-Prasad-Sommerfield (BPS) states 
whose tension is proportional to the topological charge.
These types of monopoles can be realized as kinks inside a vortex-string 
\cite{Tong:2003pz,Eto:2004rz,Eto:2006pg,Shifman:2007ce,Nitta:2010nd}.  
These monopole-vortex complexes are a story in $d=3+1$ dimensions, 
and here we extend them to confined solitons in diverse dimensions.

In $d=2+1$ dimensions, a vortex is a particle-like soliton. 
In the vortex confinement, a vortex and an anti-vortex are connected by 
a domain wall or domain line, which is string-like in this dimensionality 
[Fig.~\ref{fig:confined-defects}(a)]. 
This was used in a seminal paper by Polyakov for a proof of 
the confinement of compact QED in $d=2+1$ dimensions \cite{Polyakov:1976fu}. 
A ``quark" and ``anti-quark" in QED are mapped by a duality 
to a vortex and an anti-vortex, respectively. 
And then, a domain wall generated by a quantum effect connects them 
to result in the confinement. 
This is also promoted to 3+1 dimensions with one compactified dimension 
in an attempt to show confinement in 3+1 dimensions \cite{Shifman:2008yb}. 
Vortices connected by a domain wall are also known in several condensed matter systems  
such as multi-gap superconductors \cite{Tanaka:2001,Tanaka:2010}
and multi-component Bose-Einstein condensates 
with a Rabi coupling \cite{Son:2001td,Kasamatsu:2004}.
In the second (stable) type of vortex confinement, 
a vortex is attached by domain walls with the same tension from both of its sides 
[Fig.~\ref{fig:confined-defects}(b)] \cite{Ritz:2004mp,Eto:2005sw,Nitta:2012xq}. 
This confined vortex can be realized as a kink inside a domain wall.
In condensed matter physics, this structure appears as
a Bloch line in a Bloch wall in magnetism \cite{Chen:1977}, 
half-quantized vortices trapped in a chiral domain wall in 
chiral p-wave superconductors \cite{Garaud:2012},  
and a Mermin-Ho vortex within a domain wall in 
superfluid $^3$He (see Fig.~16.9 of Ref.~\cite{Volovik2003}). 
Also, a magnetic flux is confined as a Josephson vortex  
in a Josephson junction of two superconductors \cite{Ustinov:1998}, 
where the junction can be identified with an infinitely heavy domain wall.
These configurations can be linearly extended into $d=3+1$ dimensions, 
where vortex-strings are attached at the edges of a domain wall, that is, 
a domain wall is bounded by vortex-strings \cite{Kibble:1982dd,Vilenkin:2000}. 
In the first (unstable) type of vortex confinement, 
a domain wall is metastable and can decay quantum mechanically, 
creating a two dimensional hole bounded by a closed vortex-string \cite{Preskill:1992ck,Son:2001td}. 

\begin{figure}[ht]
\begin{center}
\begin{tabular}{cc}
\includegraphics[width=0.51\linewidth,keepaspectratio]{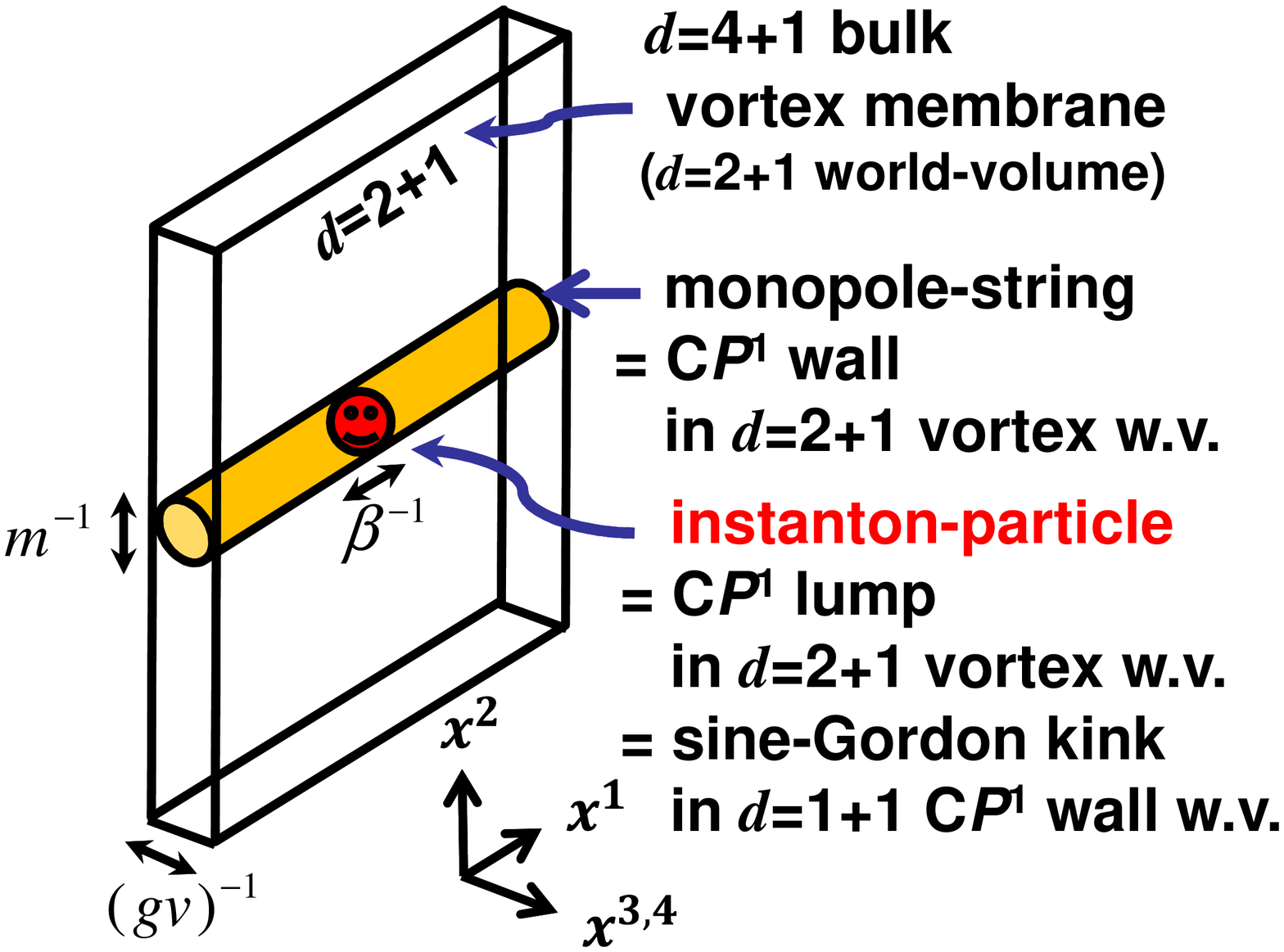}
&
\includegraphics[width=0.49\linewidth,keepaspectratio]{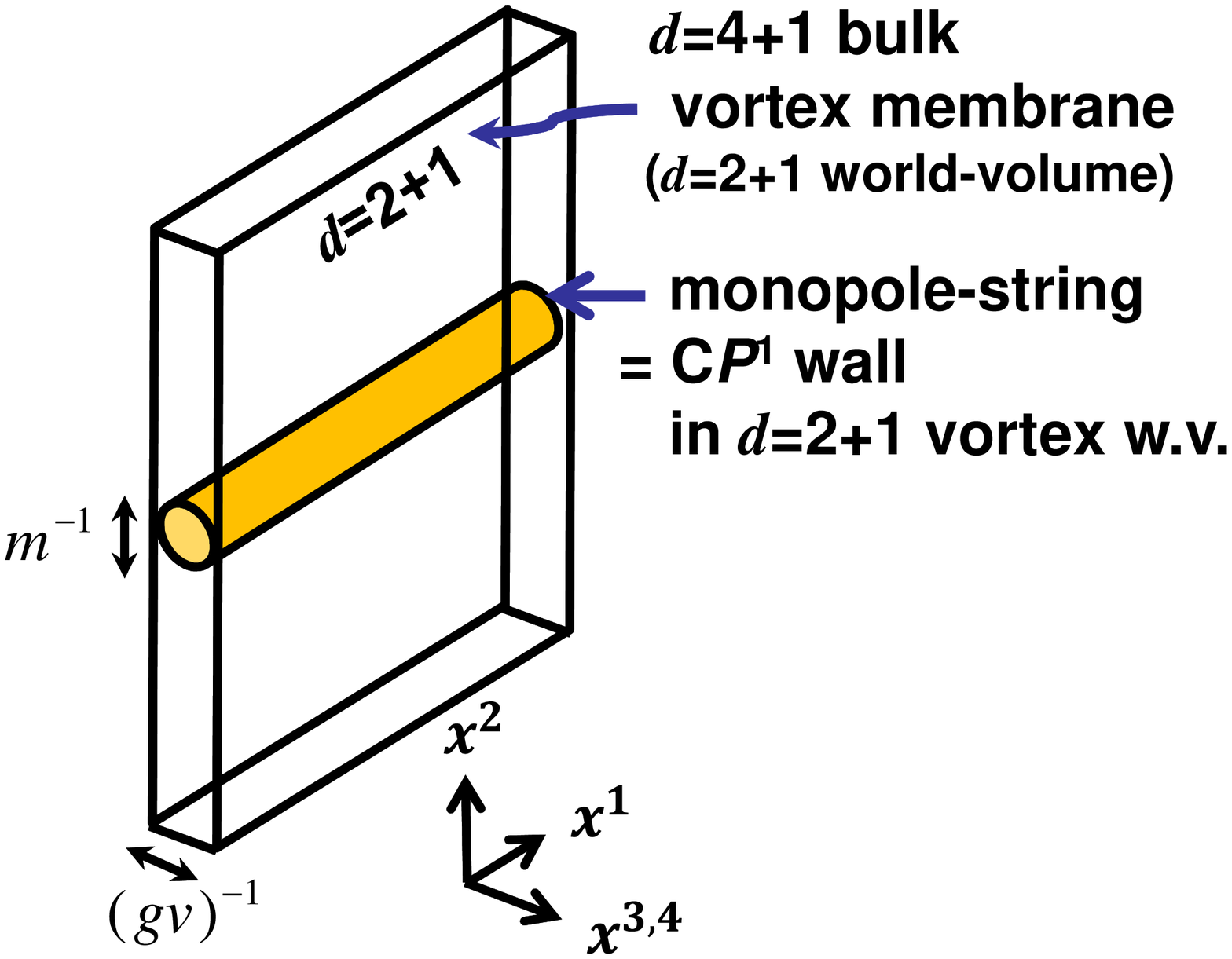}
\\
(a) & (b) \\
\includegraphics[width=0.48\linewidth,keepaspectratio]{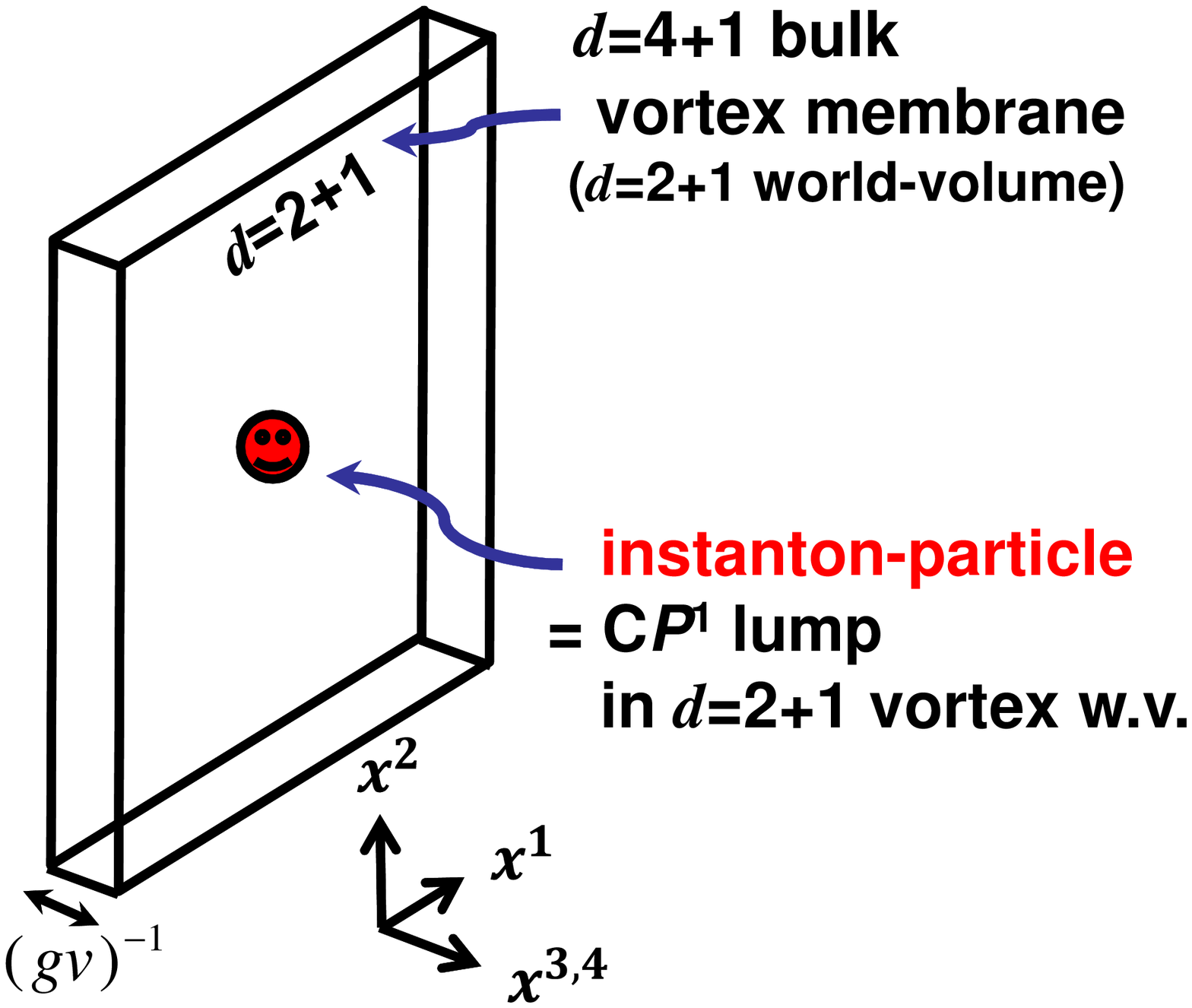}
&  
\includegraphics[width=0.49\linewidth,keepaspectratio]{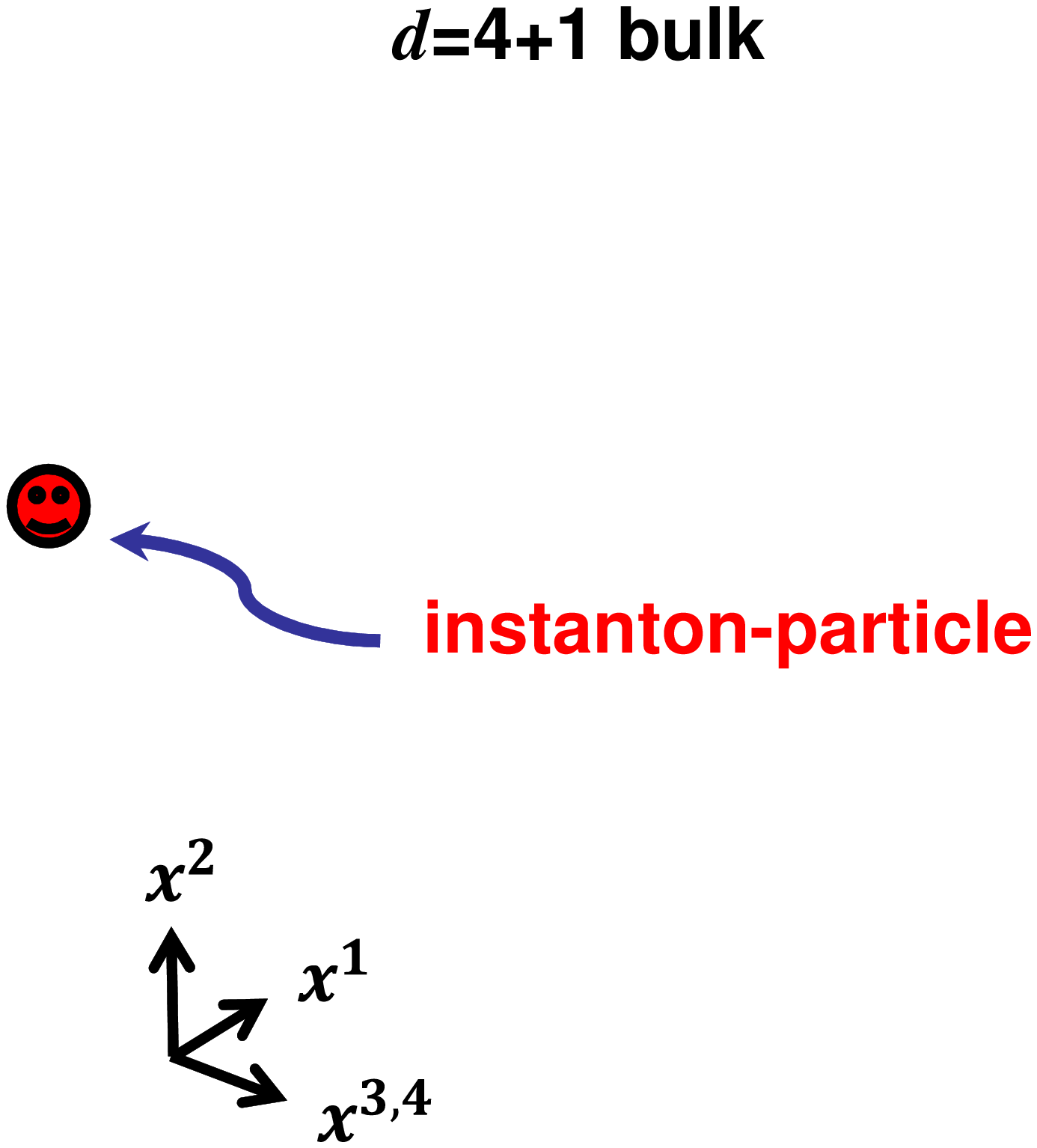}
\\
(c) & (d) \\
\end{tabular}
\end{center}
\caption{
(a) Original configuration, 
(b) $\beta \to 0$,
(c) $\beta, m \to 0$ with $\beta/m^2 = {\rm fixed}$,
(d) $\beta, v, m \to 0$ with $\beta/m^2 = {\rm fixed}$.
\label{fig:instantons}
}
\end{figure}
In this paper, we discuss a $d=4+1$ dimensional analog of these confined solitons. 
In this dimensionality,  Yang-Mills instantons are particle-like and 
't Hooft-Polyakov monopoles \cite{'tHooft:1974qc} are string-like. 
Therefore, it is natural to expect 
that a Yang-Mills instanton is confined by monopole-strings 
as in Figs.~\ref{fig:confined-defects}(e) and \ref{fig:confined-defects}(f). 
Here, we will show that this is indeed the case in a certain situation 
by focusing on the second (stable) type of confinement, 
that is, instantons can be realized as sine-Gordon kinks on 
a monopole-string.
However, it is not straightforward to show this. 
A 't Hooft-Polyakov monopole has a $U(1)$ modulus 
related to unbroken $U(1)$ gauge symmetry in the vacuum. 
This fact implies that a phase kink winding around 
the $U(1)$ modulus along the string is unstable against expansion, 
and that the kink on the string as an instanton is diluted along the string. 
In order to stabilize it, we need to introduce a potential term along the $U(1)$ modulus. 
However, since this $U(1)$ modulus comes from 
part of an $SU(2)$ gauge group, 
it is difficult to have a potential term along the $U(1)$ modulus. 
The idea to overcome this problem here is 
to put the theory into the Higgs phase 
by introducing Higgs fields in the fundamental representation with 
the vacuum expectation values (VEVs). 
Non-Abelian vortices exist in this theory \cite{Auzzi:2003fs,Hanany:2003hp,Eto:2006pg,Shifman:2007ce}, 
and a monopole-string becomes a domain wall inside a 
non-Abelian vortex \cite{Tong:2003pz,Eto:2006pg,Shifman:2007ce}.
We then introduce additional couplings among the Higgs fields, 
which we call the non-Abelian Josephson term.  
In this setup, we show that a sine-Gordon potential term 
is induced on the effective theory along the monopole-string 
and that a sine-Gordon kink on the monopole-string
as shown in Fig.~\ref{fig:instantons}(a) 
indeed carries the instanton charge.

In Sec.~\ref{sec:monopole}, we give our model, 
realize a monopole-string inside a non-Abelian vortex sheet,  
and construct the effective thoery on the monopole-string. 
In Sec.~\ref{sec:instanton}, we deform the model by a non-Abelian Josephson term, find the monopole effective Lagrangian that becomes the sine-Godron model, 
and construct instanton beads on a monopole-string as sine-Gordon kinks. 
We discuss various limits in Fig.~\ref{fig:instantons}. 
Section \ref{sec:summary} is devoted to a summary and discussion.

\section{Monopole-strings confined by vortex membranes}\label{sec:monopole}

We consider 't Hooft-Polyakov monopole-strings \cite{'tHooft:1974qc}
in the BPS limit in a $U(2)$ gauge theory 
in $d=4+1$ dimensions.
We put the system into the Higgs phase,  
where magnetic fluxes from a monopole are squeezed 
into vortices \cite{Tong:2003pz}, 
which are membranes in $d=4+1$ dimensions.
The Lagrangian which we consider in $d=5+1$ dimensions
is given by $(A,B=0,1,2,3,4)$
\beq
&& {\cal L} = - \1{4 g^2}\tr F_{AB}F^{AB} 
 + \1{2g^2} \tr (D_{A} \Sigma)^2 
  + \tr D_{A}H^\dagger D^{A}H - V ,\non
&& V = g^2 \tr (H H^\dagger -v^2{\bf 1}_2)^2 
 + \tr \left[H(\Sigma - M)^2 H^\dagger\right],
 \quad  \label{eq:Lagrangian}
\eeq
with a two by two matrix of complex scalar fields $H$ 
with $D_{A}H = \del_A H -ig A_A H$ 
and a two by two real matrix of adjoint scalar fields $\Sigma$ 
with $D_{A}\Sigma = \del_A \Sigma -ig [A_A ,\Sigma]$. 
In $d=4+1$ dimensions, this Lagrangian 
can be made ${\cal N}=2$ supersymmetric 
({\it i.e.}, with eight supercharges)
by suitably adding fermions. 
The mass matrix is given by  
$M={\rm diag.}(m_1,m_2)$, with $m_1>m_2$ and $m_1-m_2\equiv m$.
The constant $v^2$ giving a VEV to $H$ is called the Fayet-Iliopoulos parameter 
in the context of supersymmetry.
In the limit of vanishing $v^2$, 
the system goes back into the unbroken phase 
and $H$ decouples in the vacuum.
In the massless limit $m=0$, 
the Lagrangian (\ref{eq:Lagrangian}) 
enjoys the $SU(2)_{\rm F}$ flavor symmetry which acts on 
$H$ from its right. 
In this case, the system is in the so-called color-flavor locked vacuum 
$H={\rm diag.}(v,v)$ and $\Sigma=0$, in which both the color $U(2)$ and flavor $SU(2)_{\rm F}$ 
symmetries are spontaneously broken, 
with the color-flavor locked symmetry 
$SU(2)_{\rm C+F}$ remaining. 
In this case, the model admits a non-Abelian $U(2)$ vortex solution \cite{Hanany:2003hp}, 
$H = {\rm diag.}\, (f(r) e^{i\theta}, v)$, 
where $(r,\theta)$ are polar coordinates in the $x^3$-$x^4$ plane, where the vortex world-volume has the 
coordinates $(x^0,x^1,x^2)$.
The transverse width of the vortex is $1/gv$. 
The vortex solution breaks the vacuum symmetry 
$SU(2)_{\rm C+F}$ into $U(1)$ in the vicinity of the vortex, 
and consequently 
there appear ${\bf C}P^1 \simeq U(2)_{\rm C+F}/U(1)$ 
Nambu-Goldstone modes localized around the vortex. 
The vortex solutions have the orientational moduli 
 ${\bf C}P^1$ in addition to the translational (position) moduli $z$.  
By promoting the moduli to the fields depending on 
the world-volume coordinates $(x^0,x^1,x^2)$, 
the low-energy effective theory of these modes 
can be constructed to yield 
the ${\bf C}P^1$ model in  $d=2+1$ dimensions.
In the presence of mass, {\it i.e.} $m\neq 0$, 
the $SU(2)_{\rm C+F}$ symmetry is explicitly broken.
Considering a regime $m \ll gv$ of small mass, 
it induces the mass in the $d=2+1$ dimensional 
vortex effective theory \cite{Hanany:2003hp,Tong:2003pz,Eto:2004rz,Eto:2006uw}  
($\mu=0,1,2$)
\beq
&& {\cal L}_{\rm vort.eff.}
=  2 \pi v^2 |\del_{\mu} z_0|^2 + {4\pi \over g^2} \left[ 
 {\partial_{\mu} u^* \partial^{\mu} u - m^2 |u|^2 
  \over (1 + |u|^2)^2} \right] .\label{eq:vortex-th}
\eeq
Here $z_0(x^{\mu}), u(x^{\mu}) \in {\bf C}$ represent 
the position and orientational moduli 
(the projective coordinate of ${\bf C}P^1$), respectively.
The vacua in the vortex theory are $u=0$ and $u=\infty$ corresponding to 
the north and south poles of the target space 
${\bf C}P^1$. 
This is known as the massive ${\bf C}P^1$ model, 
which can be made supersymmetric with fermions 
\cite{Abraham:1992vb}.

A monopole solution can be constructed as a domain wall interpolating the two vacua 
$u=0$ and $u=\infty$ \cite{Abraham:1992vb} in the vortex effective theory (\ref{eq:vortex-th}).
We place it perpendicular to the $x^2$-coordinate as
\beq
 u_{\rm mono.}(x^2) = e^{\mp m (x^2-Y) + i \ph} , \label{eq:wall-sol}
\eeq
where $\mp$ represents a monopole and an anti-monopole 
with the width $1/m$.
Here, $Y$ and $\ph$ are moduli parameters 
representing the position in the $x^2$-coordinate and $U(1)$ phase 
of the (anti-)monopole. 
The domain wall tension 
$E_{\rm wall}= {4\pi \over g^2} \times m = E_{\rm mono.}$
coincides with the monopole mass $E_{\rm mono.}$ 
and the monopole charge in the bulk theory.
Therefore, the wall in the vortex theory 
is nothing but a monopole-string from  
the bulk point of view \cite{Tong:2003pz} 
as schematically shown in Fig.\ref{fig:instantons}(b).  
In the limit $v \to 0$ where the system goes back to an unbroken phase, 
the vortex disappears and the monopole goes back to the original 't Hooft-Polyakov monopole.

Let us construct the effective theory of the monopole-string 
by promoting the moduli $Y$ and $\ph$ to fields 
 $Y(x^i)$ and $\ph(x^i)$ ($i=1,2$) \cite{Manton:1981mp,Eto:2006uw} 
on the string: 
\beq
 {\cal L}_{\rm mono.eff.} 
&=& {4\pi \over g^2} \int_{-\infty}^{+\infty} dy
 {e^{2my} \over (1+e^{2my})^2} [(\del_i Y)^2 + (\del_i\ph)^2]\non
&=& {4\pi \over g^2} \1{2m} [(\del_i Y)^2 + (\del_i\ph)^2].
\eeq
We now have a free theory, a sigma model with the target space 
${\bf R}\times U(1)$. 
As denoted in the introduction, 
even if we consider a phase kink winding around the 
$U(1)$ modulus along the string, it is unstable against expansion and is diluted along the string. 
In order to stabilize it, we need a potential term 
which prevents the expansion.

\section{Instanton beads on a monopole-string}\label{sec:instanton}
In order to stabilize a phase kink on the monopole-string, 
we introduce the following deformation term in the original Lagrangian in $d=4+1$:
\beq
 \Delta {\cal L}  =  - {\beta^2 \over v^2} \tr (H \sigma_x H^\dagger ). \label{eq:deform}
\eeq
We call this term a ``non-Abelian Josephson term"\footnote{
The term  in Eq.~(\ref{eq:Josephson}) can be rewritten as 
$c v^2\beta^2 \phi^{*1}\phi^2$ 
in terms of homogeneous coordinates 
$(\phi^1,\phi^2)$ of ${\bf C}P^1$ with the identification
$(\phi^1,\phi^2) \sim e^{i\alpha}(\phi^1,\phi^2)$ and 
the constraint $|\phi^1|^2+|\phi^2|^2=1$.
This term is a Josephson term 
appearing in a Josephson junction of two superconductors 
with two condensates $\phi^1$ and $\phi^2$. 
In this sense, the term in Eq.~(\ref{eq:deform}) can be called 
the non-Abelian Josephson term.
}.
We work in the parameter region $\beta \ll m v$ to treat this term perturbatively and assume the vortex solution is not modified at the leading order. 
Then, this term induces a potential term in the vortex effective Lagrangian 
\cite{Eto:2009tr}:
\beq
\Delta {\cal L}_{\rm vort.eff.} =
 - c \beta^2 D_x .  \label{eq:Josephson}
\eeq
Here, $D_x$ is a so-called moment map, defined by
\beq
 D_x = {u + u^*\over 1+|u|^2} ,
\eeq
and $c$ is a numerical constant given by the integration
\beq
 c = \sqrt{2} \pi \int_{0}^{\infty} dr\, r (v^2 - f^2 ) \equiv {\tilde c \over g^2} 
(>0), 
\eeq
with the vortex profile function $f$.
In order to evaluate this integral, we approximate
the profile function as 
$f=gv^2r$ for $r \leq 1/gv$ and $f=v$ for $r\geq 1/gv$.
We then get:
\beq
 \tilde c  \sim {\sqrt{2}\pi \over 4} \sim {1.11} .
 \label{eq:c}
\eeq

Since we are working in the parameter region of small $\beta$, 
we can assume that the monopole solution (\ref{eq:wall-sol}) is not deformed 
at the leading order. 
Then, the monopole effective Lagrangian is deformed by
\beq
 \Delta {\cal L}_{\rm mono.eff.} 
 = c \beta^2 \int_{-\infty}^{+\infty} dy 
 {e^{my+ i\ph} + e^{my- i\ph} \over 1+e^{2my}} 
 = {\pi c \beta^2 \over m} \cos \ph.
\eeq
Finally we arrive at the monopole effective Lagrangian summarized as
\beq
 {\cal L}_{\rm mono.eff.} +  \Delta {\cal L}_{\rm mono.eff.} 
 &=& {4\pi \over g^2} \1{2m} [(\del_i Y)^2 + (\del_i\ph)^2 ] 
+ {\pi c \beta^2 \over m} 
\cos \ph \non
&=& {2\pi \over g^2 m} [(\del_i Y)^2 + (\del_i\ph)^2  
+ \tilde \beta^2 
\cos \ph], \label{eq:SG}
\eeq
with 
\beq
 \tilde \beta^2 \equiv \1{2} {\tilde c }\beta^2 \sim 
 0.555 \beta^2.
\eeq
Here we have used the estimation of $c$ in Eq.~(\ref{eq:c}). 
The monopole effective Lagrangian in Eq.~(\ref{eq:SG}) is 
the one of the sine-Gordon model with the additional field $Y$.

Let us construct a sine-Gordon kink. 
The Bogomol'nyi completion for the energy density is 
\beq
 {g^2 m \over 2 \pi} E &=& (\del_i\ph)^2 + \tilde \beta^2 \left(\sin^2 {\ph\over 2} -1\right) \non
 &=& \left(\del_i\ph \pm \tilde \beta \sin {\ph\over 2}\right)^2 
  \mp 2 \tilde \beta \partial_i \ph \sin {\ph\over 2} - \tilde \beta^2\non
 &\geq& {g^2 m \over 2 \pi} |t_{\rm SG}| - \tilde \beta^2
\eeq
with the topological charge density
\beq
t_{\rm SG} 
 \equiv { \tilde \beta \over m} \partial_i \ph \sin {\ph\over 2}
 = - { 2 \tilde \beta \over m} \partial_i \left( \cos {\ph\over 2}\right).
\eeq
The inequality is saturated by the BPS equation
\beq
 \del_i\ph \pm \tilde \beta \sin {\ph\over 2} = 0.
\eeq
For instance, the one-kink solution and its topological charge 
are 
\beq
 \ph = 4 \arctan \exp{{\tilde \beta \over 4} (x- X)} 
 + {\pi \over 2};\ \quad
 T_{\rm SG} = \int  dy t_{\rm SG} = {4\tilde \beta \over m}.
\eeq
The width of the kink is $\Delta x \sim 1/\tilde \beta$.
The energy of the one sine-Gordon kink is 
\beq
 E_{\rm SG} =   {2 \pi \over g^2 m} T_{\rm SG} 
= {8 \pi \tilde \beta \over g^2 m^2} 
= 4 \sqrt 2 \pi \sqrt{\tilde c} {\beta \over g^2 m^2} 
 \sim 18.7 {\beta \over g^2 m^2} .
  \label{eq:SG-energy}
\eeq

What does the sine-Gordon kink correspond to in the vortex theory 
and in the bulk?
First, one can confirm that 
 $k$ sine-Gordon kinks can be identified with $k$ ${\bf C}P^1$-lumps with  
the topological charge $k \in \pi_2 ({\bf C}P^1)$ \cite{Polyakov:1975yp} 
in $d=2+1$ dimensional vortex world-volume,  
by explicitly calculating a lump charge
\beq
 T_{\rm lump} 
&\equiv& \int d^2x {i (\partial_i u^* \partial_j u - \partial_j u^* \partial_i u )
\over (1+|u|^2)^2} \non
&=& \oint dx^i {-i (u^* \partial_i u - (\partial_i u^*) u )
             \over 2 (1+|u|^2)} \non
&=& \oint dx^i {|u|^2  \over 1+|u|^2} \partial_i \ph
= 2 \pi k .
\eeq

The ${\bf C}P^1$ lumps 
in the vortex effective theory 
can be further identified with Yang-Mills instantons in the bulk \cite{Eto:2004rz} 
as can be inferred from the lump energy $E_{\rm lump} $, 
coinciding with the instanton energy $E_{\rm inst}$ \cite{Eto:2004rz}:
 $E_{\rm lump} = {4\pi \over g^2} T_{\rm lump} 
= {4\pi \over g^2} \times 2\pi k 
= {8\pi^2 \over g^2} k = E_{\rm inst}$. 
Therefore, we conclude that the sine-Gordon kink
on the monopole-string 
corresponds to a Yang-Mills instanton in the bulk point of view,   
as schematically shown in Fig.\ref{fig:instantons}(a).
Multiple sine-Gordon kinks give multiple instanton beads 
on the monopole-string.

To clarify the relations with configurations known before, 
let us discuss various limits shown in Fig.~\ref{fig:instantons}.

\begin{enumerate}
\item
$\beta \to 0$ [Fig.~\ref{fig:instantons}(b)]: 
Monopole-string confined by vortex-sheets \cite{Tong:2003pz}. \\
The instanton is diluted and eventually disappears because
the energy of the sine-Gordon kink becomes zero.

\item \label{limit3}
$\beta, m \to 0$ with $\beta/m^2 ={\rm fixed} = {\sqrt 2 \pi \over \sqrt {\tilde c}}$ [Fig.\ref{fig:instantons}(c)]: Instanton particle trapped inside a vortex sheet \cite{Eto:2004rz}. \\
The monopole-string disappears while the vortex-sheet and instanton remain.
The scaling is determined by requiring the energy of a sine-Gordon kink 
in Eq.~(\ref{eq:SG-energy})
to reduce to the BPS instanton energy $= 8\pi^2/g^2$.  
The precise value of the constant $\tilde c$ is needed 
to determine the scaling constant.

\item
$\beta, v, m \to 0$ with $\beta/m^2 = {\rm fixed}  = {\sqrt 2 \pi \over \sqrt {\tilde c}}$ [Fig.\ref{fig:instantons}(d)]:
Bare instanton particle.\\
Only the instanton remains while the vortex-sheet and monopole-string disappear. The transverse size of the vortex $1/gv$ grows,  
and the vortex is diluted and disappears from the limit \ref{limit3}. 
\end{enumerate}

From these limits, one may think that
the simple limit $v \to 0$ gives
an instanton on a monopole-string 
without a vortex sheet. 
However, the vacuum expectation value of $H$ vanishes in the limit $v \to 0$ 
so that the deformation by $\beta$ 
in Eq.~(\ref{eq:deform}) does not seem to 
affect the configuration. 
The instanton will be diluted along 
the infinitely long monopole-string. 
This limit seems to be rather subtle. 

In this paper, we have constructed the stable type of instanton confinement
in Fig.~\ref{fig:confined-defects}(f). 
In a model where a monopole-string is unstable, 
the monopole-string decays through quantum tunneling by  
creating an instanton and an anti-instanton as 
shown in Fig.~\ref{fig:confined-defects}(e).

\section{Summary and Discussion}\label{sec:summary}
We have constructed instanton beads on a monopole-string 
in $d=4+1$, which shows that an instanton is confined by monopole-strings 
as a higher dimensional generalization of a monopole confined by vortex-strings in $d=3+1$ and a vortex confined by domain-lines 
in $d=2+1$. 
We have constructed this configuration inside a non-Abelian vortex sheet 
by putting the system into the Higgs phase 
in supersymmetric gauge theory with a deformation term, 
which we call the non-Abelian Josephson term. 
A monopole-string has been realized as a domain-line 
inside the vortex,  
and we have found that 
the effective theory on the monopole-string 
is the sine-Gordon model whose potential is induced 
by the non-Abelian Josephson term.
Then, the instanton has been realized 
as a sine-Gordon kink in the monopole effective theory.
We have discussed several limits of the configuration reducing to 
previously known configurations such as 
a confined monopole on the vortex 
and an instanton trapped inside the vortex.

A necklace of instantons can be constructed 
by making a closed monopole-string inside a non-Abelian vortex. 

It is unclear thus far whether instanton beads on a monopole-string can exist without the help of a vortex. 
Without the potential on the monopole theory, 
the instanton charge made of phase kinks will be diluted along 
the infinitely long monopole-string and disappear.  
However, if one makes phase kinks on a closed monopole-string, 
one gets a twisted closed monopole-string where the instanton charge 
is uniformly distributed along the loop. 
It is nothing but an instanton \cite{Nitta:2012kj}.  

We have studied $U(2)$ gauge theory with two flavors, 
while a generalization to $U(N)$ gauge group with $N$ flavors 
can be made.
When we consider non-degenerate masses for $H$, 
there appear $N-1$ parallel monopole-strings 
inside the vortex-sheet as $N-1$ domain walls \cite{Isozumi:2004jc}. 
By considering a non-Abelian Josephson term such as
$\sum_{a \in {\rm root}} \beta_a^2 \tr (HT_a H^\dagger)$,
we would have sine-Gordon kinks in multiple monopole-strings, 
which remain as a future problem. 
Moreover, if we consider 
degenerate masses for $H$ in the $U(N)$ case,  
monopoles become non-Abelian monopoles \cite{Nitta:2010nd} inside a vortex 
as non-Abelian domain walls \cite{Shifman:2003uh,Eto:2005cc,Eto:2008dm}. 
Then, a non-Abelian generalization of sine-Gordon kinks should be considered as instantons on a non-Abelian monopole-string.
Generalization to arbitrary gauge groups 
is also possible as was so for a non-Abelian vortex \cite{Eto:2008yi}.

\section*{Acknowledgements}

This work is supported in part by 
a Grant-in-Aid for Scientific Research (No. 23740198) 
and by the ``Topological Quantum Phenomena'' 
Grant-in-Aid for Scientific Research 
on Innovative Areas (No. 23103515)  
from the Ministry of Education, Culture, Sports, Science and Technology 
(MEXT) of Japan. 


\end{document}